\documentclass[aps,prl,twocolumn,groupedaddress]{revtex4}
\usepackage{graphicx}
\begin{document}


\title{Effective Mass Anomalies in Strained Si Thin Films and Crystals}


\author{Jun Yamauchi}
\affiliation{
Faculty of Science \& Technology, Keio University, 3-14-1 Hiyoshi, Yokohama, 223-8522, Japan\\
Institute of Industrial Science, University of Tokyo, 4-6-1 Komaba,
 Meguro-ku, Tokyo 153-8580, Japan  \\
}


\date{\today}

\begin{abstract}
 Effective mass anomalies due to the geometrical effects are
 investigated in silicon nanostructures using first-principles
 calculations for the first time.
 In \{111\} and \{110\} biaxially strained Si, it is found 
 that longitudinal effective mass is extraordinarily enhanced
 for both thin films and crystals.
 This mass enhancement is caused by the change of 
 the band structure with double minima into that
 with a single minimum due to strain and confinement.
 At the transition point, it is analytically shown that the effective
 mass diverges.
 The dependences of the confinement thickness on the anomalies are
 qualitatively explained by an extension of the effective mass approximation.
\end{abstract}


\maketitle

\section{}
\subsection{}
\subsubsection{}

In this decade, various  nanostructures have been
fabricated and measured. In such small structures,
the geometrical effect due to strain, confinement, and crystal directions
significantly changes the electronic structures compared with that of the
 bulk.
  Intentional use of the geometrical effect
  has been suggested for the improvement of the device
  performance\cite{Takagi96}.
  From the technological viewpoint,
  the function of sub-nm-body metal oxide field-effect
  transistors (MOSFETs) using
  silicon on insulator\cite{SOI} has been demonstrated,
  which consist of a silicon
  transport region sandwiched by insulating layers\cite{Uchida03}.
  The nanostructure most sensitive to the geometrical effect
  may be an 
  ultrathin body double-gate (UTBD) MOSFET\cite{Balestra87,Gamiz01}
  with  a body thickness of less than 10nm;
  it has a three dimensional
  structure with the channel region sandwiched by two gate insulators
  and is a promising sub-10 nm device.
  Recently, in terms of electron mobility for the UTBD MOSFET,
  the advantage of the $\langle 110\rangle$ direction due to the valley structure
  has been reported in comparison with the $\langle 001\rangle$ direction\cite{Tsutsui05}.
  It is very interesting to investigate 
  the change of the electronic structures including
  transport parameters such as effective mass
  from the bulk to the thin films of some atomic layers.
  However, there are few studies on the intrinsic properties of
  semiconductor films 
  in the intermediate region from the bulk to the sub-nm body using
  first-principles calculations\cite{Yamauchi05}.

  In this Letter, 
  the effective mass is systematically investigated
  for the biaxially strained Si thin films
  confined in the $\langle 001\rangle $, $\langle 111\rangle$, and
  $\langle110\rangle$ directions and for 
  strained Si crystals  using  density functional calculations, where
  the strain is uniformly applied to the plane perpendicular to the
  confinement direction.
  In the \{111\} and \{110\} strained Si, it is found 
  that in some region, longitudinal effective mass is extraordinarily enhanced
  for the thin films and the crystals.
  Preliminary results are reported in \cite{Yamauchi07}.
  The confinement and the symmetry breaking due to the strain
  cause these anomalies, where the band structures change
  from double minima to a single minimum.
  In such a case, it is analytically shown that the effective mass
  diverges. 
  As far as the author knows, this is the first
  report on the qualitative change of the band structure and the
  divergence of the effective mass in Si caused 
  by confinement and strain\cite{Rideau}. 
  The dependences of the confinement thickness on the anomalies are
  qualitatively explained by extending the effective mass approximation
  to the whole band near the Brillouin zone boundary.
  The experimentally observable regions of the strain and thickness
  are also obtained for the $\langle110\rangle$ confinement.

 Band calculation is based on the density functional
 theory\cite{Hohenberg64,Kohn65}.
 The local density approximation (LDA) PW92\cite{Perdew92}
 is adopted for the exchange correlation functional.
 To describe ion-electron interaction, norm-conserving\cite{Troullier90}
 and ultrasoft\cite{Vanderbilt90} pseudopotentials are adopted for  Si and Cl,
 and for H, respectively.
 The calculational condition is carefully checked\cite{calc-cond}.
 The cutoff energy is 20.25 Rydberg.
 The meshes of the sampled k-points are $8\times 8$, $8\times 8$,
 and $4\times 8$ for the (001), (111) and (110) slabs, respectively.
 The calculated longitudinal and transverse masses for the bulk without
 strain are 0.950 (0.9163) and 0.189 (0.1905), respectively.
 The figures in the
 parentheses are experimental ones\cite{LB}.
 The unit of effective mass is the bare electron mass.
  The effective mass is calculated from a
  fitting of eigenvalues  to the 6-th-order polynomial.
 The calculation is performed using 
 Tokyo Ab initio Program Packages (TAPP) code \cite{TAPP}.

 As a model of Si film sandwiched by insulating layers, 
 hydrogen-terminated Si $(001)$, $(111)$  and  
 $(110)$-($1\times 1$) slab models are used for  the [001], [111] and
 [110] confinements, respectively.
 The plane perpendicular to the confinement direction is uniformly
 strained from --4 to +4\%\cite{lackofdata}
 and the lattice constant along the confinement direction is relaxed
\cite{internal-parameter}.
 The slab models with hydrogen and chlorine terminated for each side
 are additionally used to evaluate the surface effect.
 The Si atoms in the model consist of the crystal structure unit,
 which is optimized for the given strain. 

\begin{figure}
  \includegraphics[scale=0.35]{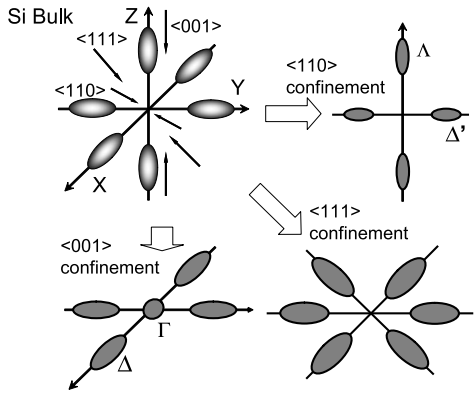}
  \caption{\label{fig:e-pockets}
  Schematic illustrations of the electron pockets for the bulk
  and the confined thin films {\em without} strain. For the bulk, six
  equivalent ellipsoidal electron pockets are located along the
  $\Gamma-X$ lines. 
  Confinement projects these electron pockets onto the perpendicular
  plane: for $\langle001\rangle$, one isotropic valley ($\Gamma$-valley) and
  four elliptic valleys ($\Delta$-valleys); for $\langle110\rangle$ ,two kinds of
  inequivalent valleys ($\Lambda$- and $\Delta'$-valleys), where the two and
  four pockets in the bulk correspond 
  to the $\Lambda$- and the $\Delta '$-valleys, respectively;
  for $\langle111\rangle$, six equivalent elliptic pockets. 
   }
 \end{figure}

 Before proceeding to the results, the electron pockets for the 
 confined thin films are briefly introduced. 
 For the {\em non-strained} case, Fig.~\ref{fig:e-pockets}
 shows schematic illustrations of the electron pockets of the bulk
 and the slabs confined in the $\langle001\rangle$, $\langle110\rangle$ and $\langle111\rangle$ directions.
 For the $\langle001\rangle$- and $\langle110\rangle$-confined slabs,
 there are two kinds of inequivalent valleys.
 The strain and the confinement thickness
 determine which valley is the bottom of the conduction band.
 Among them, the $\Gamma$- and the $\Lambda$-valleys are mainly focused. 

\begin{figure}
  \includegraphics[scale=0.37]{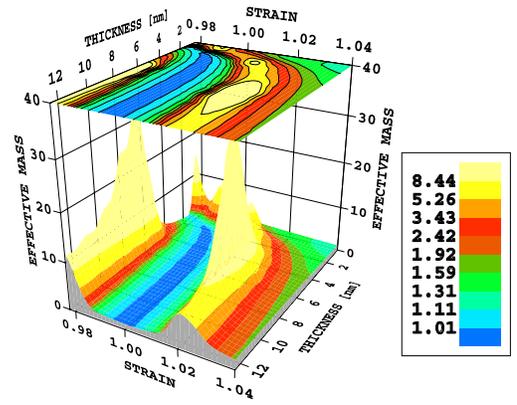}
  \caption{\label{fig:emass110}
  Longitudinal effective mass at the $\Lambda$-valley (Fig.~\ref{fig:e-pockets})
 in the $\langle110\rangle$-confined slab as a function of the strain and
  thickness. 
  The unit of effective mass is the bare electron mass. The strain is
  represented by the ratio of the strained lattice constant to
  the bulk value without strain.
  The strain more/less than 1 corresponds to the tensile/compressive
  strain applied  biaxially to the plane perpendicular to the confinement
  direction.
  For the effective mass, values more than 50 are truncated to avoid
  spurious interpolation. 
}
 \end{figure}

\begin{figure}
  \includegraphics[scale=0.36]{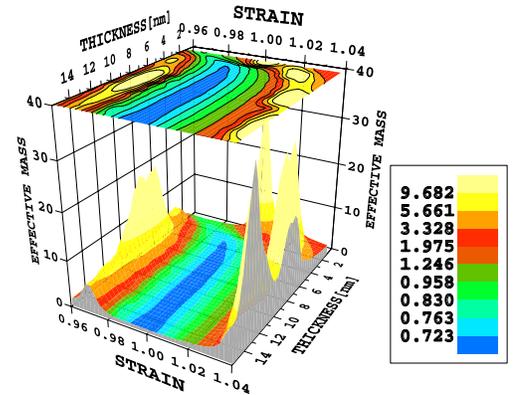}
  \caption{\label{fig:emass111}
  Longitudinal effective mass  in the [111]-confined slab
  as a function of the strain and thickness. 
  The configuration of the graph is the same as in Fig.~\ref{fig:emass110}.
  }
 \end{figure}

\begin{figure}
  \includegraphics[scale=0.35]{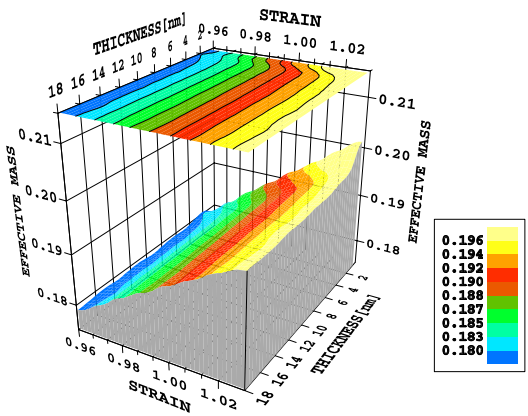}
  \caption{\label{fig:emass100}
  Effective mass at the $\Gamma$-valley (Fig.~\ref{fig:e-pockets})
 in the [001]-confined slab
  as a function of the strain and thickness. 
  The graph configuration is the same as in Fig.~\ref{fig:emass110}.
  }
 \end{figure}

 Fig.~\ref{fig:emass110} shows the longitudinal effective mass at the
 $\Lambda$-valley in the [110] confinement as a function of the strain
 and the slab thickness.
 The strain is represented by the ratio of the lattice constant under
 strain to that for the
 strain-free bulk. The unit of thickness is nm.
 This figure clearly reveals that  along lines, there are
 two regions with extraordinarily large masses.
 The confinement causes the contour lines
 to curve toward the direction of decreaseing strain. 
 This tendency becomes stronger as the thickness decreases.
 For the [111] confinement, the longitudinal mass distribution is
 shown in Fig.~\ref{fig:emass111}. 
 On the other hand, in the [001]-confined slabs,
 the behavior of the effective mass at the $\Gamma$-valley is very
 smooth  and the variation range of the mass is narrow (0.18 $\sim$
 0.20) in Fig.~\ref{fig:emass100}.
 In particular, the effect of confinement is negligibly small.
 Note that, for the [111] and [110] confinements,
 a huge mass enhancement is observed in the strain-free region 
 as well as the strained region.
 To evaluate the surface termination effect, the slab models with the H
 and Cl atoms for each surface side are investigated and it is found
 that these models also yield qualitatively the same results as above.

\begin{figure}
  \includegraphics[scale=0.40]{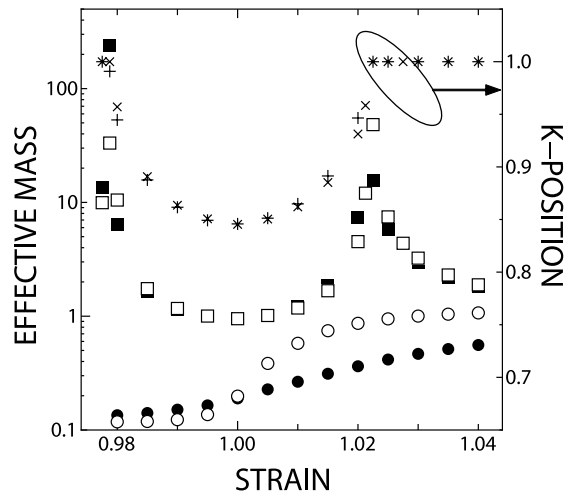}
  \caption{\label{fig:110mass-k-pos}
  Effective mass at $\Lambda$-valley 
  and minimum k-position of the $(110)$ strained
  Si thin film and bulk. 
  Longitudinal effective masses of Si slab with 65 Si layers ($\sim$ 12nm)
  (closed square)  and of Si bulk (open square).
  The k-position where the lowest conduction band takes the local
  minimum along the $\Gamma$-$X$ line ([001] direction)
  for Si slab (``+'') and for bulk (``$\times$'').
  The scale of the k-position is represented by the fractional
  coordinates, whose unit is $\Gamma X$ (half of the nearest neighbor
  reciprocal lattice vector along [001]). The Brillouin zone (BZ) of the
  slab is folded and the $X$-point at the BZ boundary
  of the bulk corresponds to the  $\Gamma$-point in the slab.
  To allow it to be drawn in the same figure, the  k-position for the slab is
  transformed by $1-\kappa/2$, where $\kappa$ is the k-position for
  the  slab.
  The {\em bulk} effective mass along the confinement direction ([110])
  at the k-positions of 0.84 (closed circle) and 0.95 (open circle).
  }
 \end{figure}

 It is reported that the behavior of effective mass
 for the slab with a thickness more than 10 nm
 is very similar to that for the bulk\cite{Yamauchi05}. 
 Fig.~\ref{fig:110mass-k-pos} shows
 a comparison between the longitudinal 
 effective mass at the $\Lambda$-valley for a slab with 65 Si atomic
 layer ($\sim$12 nm) (closed square) and for the bulk (open square)
 in the $(110)$ strained case.
 As expected, both behaviors are very similar and reveal
 a mass divergence at the strains of 0.978 and 1.022.
 At these strains, the band structure qualitatively changes
 from double minima to an single minimum.
 In Fig.~\ref{fig:110mass-k-pos}, the k-position of the minima is plotted
 for the slab (``+'') and the bulk (``$\times$''), where the position is
 represented by the fractional coordinates and  the value 1 corresponds
 to the Brillouin zone (BZ) boundary.
 Therefore, the number of electron pockets for the strongly strained bulk
 changes from six to five.
 The same analysis is valid for the $(111)$ strained
 case\cite{Yamauchi07} and the number of electron pockets changes
 from six to three.
 When the band with double
 minima changes into that with a single minimum as a function of one
 parameter, it can be analytically shown,
 using the same treatment as Landau's phenomenological 
 theory for second phase transitions,
 that the curvature at the minimum
 becomes zero; that is, the effective mass along the merging direction
 diverges.
 Considering the time reversal symmetry, the band with double minima
 near the BZ boundary along the reciprocal lattice vector
 is expressed to the lowest order of $k-k_0$ as
$\varepsilon(k,\lambda)=
 a(\lambda)+\frac{b(\lambda)}{2}(k-k_0)^2+\frac{c(\lambda)}{4}(k-k_0)^4, $
where $\lambda$ is the parameter used to change the band
structure, $k$ is the position in the BZ along the vector, and
$k_0$ is at the BZ boundary. The k points with the extrema are
given by 
$k=k_0,k_0\pm \sqrt{\frac{b(\lambda)}{c(\lambda)}}$.
Assuming that, when $\lambda \rightarrow \lambda _0$, the double
 minima merge into one, the following condition is obtained:
$\left. \sqrt{\frac{b(\lambda)}{c(\lambda)}}
\right|_{\lambda \rightarrow \lambda_0}=0$, that is,
$\left.b(\lambda)\right|_{\lambda \rightarrow \lambda_0}=0$.
This means that the second derivatives of the band energy 
vanish and the effective mass diverges at $\lambda_0$.

The mass divergence occurs  when the band structure with the double
minima changes into that with the single minimum.
This band structure change is derived from
 the strain and confinement.

 About the strain for the bulk, it is reported by  Ma {\em et al.}
 that the degeneracy of the lowest conduction bands at
 the $X$ point splits due to the (110) and (111) strains
 but does not do so due to the (001) strain
 as determined using a semiempirical tight-binding method and group theory
 \cite{Ma91,Ma93}.
 In this study, it is shown that sufficiently strong (111) and (110)
 strains largely break the symmetry, split the degenerated bands at
 the BZ boundary, and make the lower band with the single minimum at the
 BZ boundary, which causes the effective mass anomalies for the thin films
 and the bulk.

 The effect of the confinement on the effective mass
 anomalies is
 qualitatively explained  as follows. According to the
 effective mass approximation (EMA), the effect of confinement on the band
 structure is described by the increase of zeropoint energy
 $\frac{\hbar^2\pi^2}{2m_{\tt conf}}\frac{1}{L^2}$, where
$m_{\tt conf}$ and $L$ are the effective mass along the confinement
direction and the thickness, respectively.
We extend this EMA expression using the effective mass at the energy
minima to the whole band near the BZ boundary.
In Fig.~\ref{fig:110mass-k-pos}, the effective mass along the
confinement direction [110] at the k-points of 0.84 and 0.95
 in the fractional coordinates is plotted
using closed and open circles, respectively. 
The value 0.84 corresponds to the local minima without strain and
0.95 to the neighborhood of the BZ boundary.
In the tensile strain region($>1$), the mass at 0.95 is larger than that
at 0.84. Therefore, as the confinement thickness decreases, the
energy level at the 0.84 increases compared with that at the BZ boundary
and the structure with the single minimum is preferable to that with the
double minima, which well explains the behavior of effective mass
anomalies in Fig.~\ref{fig:emass110}. In the compressive region, the same
discussion is valid.
A similar discussion is found to hold for the $\langle111\rangle$ confinement case.

For the experimental observation, it is in the $\langle110\rangle$
confinement case that the effective mass anomalies have the largest
effect on the electron transport,
because, in the $\langle111\rangle$ confinement, the anomalies are averaged out by
the equivalent valleys along three different directions.
For the $\langle110\rangle$ confinement, there are two kinds of valleys:
$\Lambda$ and $\Delta'$, in which the latter does not reveal the anomalies.
The parameter region suitable for the observation
corresponds to the anomalies where the $\Lambda$-valley
 is the conduction bottom.
In the $\langle110\rangle$ confinement, there are two such regions:
that of the thickness range from the bulk to 6 nm and around the 0.98 strain, and
that of the thickness range less than 2 nm and the strain range from 0.99 to
1.01. 
It is experimentally reported that the mobility for channel
thicknesses less than 5 nm decreases
in the UTBD FET with the $\langle110\rangle$-confined\cite{Tsutsui05}.
This mobility degradation may be due to the effective mass
enhancement as well as the surface roughness scattering.
When the conduction current is large,
the anomaly points where the $\Lambda$-valley is energetically higher
than the $\Delta'$-valley may affect the
transport, because the energy difference between the $\Lambda$- and
$\Delta'$-valleys is small ($< $50 meV).
For applications that involve, combining piezoelectric devices, the effective mass
may be controlled by voltage. 

Note that, considering the mechanism of symmetry
breaking and confinement, the effective mass
anomalies are expected to be found for materials with a similar band
structure to Si such as diamond (C), SiC, AlSb, and GaP.

In summary, the effective mass of strained Si thin films and
crystals was systematically investigated using first-principles
calculations. In $(111)$ and $(110)$ strained Si, effective
mass anomalies were found for both the thin films and the crystals.
The confinement and the symmetry breaking due to the strain
changed the band structure from double minima to a single
minimum, which caused the anomalies.
In such a case, it was analytically shown that the effective mass
diverges.
The dependences of the confinement on the anomalies were
qualitatively explained by extending the effective mass approximation
to the whole band near the BZ boundary.
The experimentally observable regions of the strain and thickness
 were obtained for the $\langle110\rangle$ confinement.

The author would like to thank Dr.~S.~Matsuno and 
Professor K.~Shiraishi for valuable discussions.
This work was partially supported by a Grant-in-Aid for Scientific
Research (Nos. 17064002 and 18063003) from MEXT. Part of this research was
carried out through the  RSS21 project supported by the Research and Development for
Next-generation Information Technology of MEXT. The computations were
performed at the Research Center for Computational Science, Okazaki, Japan.


\begin{thebibliography}{99}
\bibitem{calc-cond} The cutoff energy convergence is checked up to 49
	Rydberg with respect to the effective mass at the $\Gamma$ point for
	the $\langle001\rangle$-confined 5-Si-layer slab model.
	The deviation of the effective mass is within 1\%.
	The convergence of the sampled k-points is evaluated for the response
	to the biaxial strain in comparison with the calculated values
	using the doubled mesh for each direction.
	The relative deviation is within 1\%
	for all confinement directions. The effective mass is less
	sensitive.
\bibitem{internal-parameter} In the case of the strained (111) and
	(110), it is necessary to optimize the internal coordinates,
	which is described by the internal displacement parameter
	$\zeta$. The calculated $\zeta$ is 0.54 for the both directions,
	which  shows excellent agreement with an other theoretical
	value 0.53\cite{Nielsen85}.
\end{thebibliography}
\end{document}